\documentclass[a4paper,12pt]{article}
\pdfoutput=1 

\usepackage{jcappub} 

\usepackage[T1]{fontenc} 

 \usepackage{bm}
 \usepackage{indentfirst}
 \usepackage{amsmath}
 \usepackage{graphicx}
 \usepackage{float}
 \usepackage{amssymb}
 \usepackage{subfigure}
 \usepackage{amssymb}
 \usepackage{psfrag}
 \usepackage{hyperref}
 \usepackage{epstopdf}
 \usepackage{diagbox}

\title{Inflationary attractors from a non-canonical kinetic term}

\author{Zhu Yi,}
\author[1]{and Zong-Hong Zhu\note{Corresponding author.}}

\affiliation{Department of Astronomy, Beijing Normal University, Beijing 100875, China}

\emailAdd{yz@bnu.edu.cn}
\emailAdd{zhuzh@bnu.edu.cn}

\abstract{We show explicitly how the T-model, E-model, and Hilltop inflations are
obtained from the inflation models with a non-canonical kinetic term and an arbitrary potential. By this method, any  attractor of observables $n_s$ and $r$  is possible.
The presence of attractors poses a challenge to differentiate inflation models.}

\begin{document}
\maketitle
\flushbottom

\section{Introduction}
Inflation is  a graceful solution to the   monopole, horizon, and flatness  problems in  the standard big bang cosmology; besides,  the quantum fluctuations of the inflaton can provide  the initial condition   for the formation of the large-scale structure of our Universe \cite{Guth:1980zm,Linde:1981mu,Albrecht:1982wi,Starobinsky:1980te,Sato:1980yn}.
The constraints on the inflation from the   cosmic microwave  background (CMB)
anisotropies are $n_s=0.9649\pm 0.0042$ (68\% C.L.) and $r_{0.002}<0.064$ (95\% C.L.)  \cite{Akrami:2018odb}.
For easy comparison with observational data,  the predictions of the inflation models  are generally expressed in terms  of  the number of $e$-folds $N$ at a pivot scale before the end of inflation.
According to the  observational data, the simplest form of the spectral index may be  $n_s=1-2/N$  with  the $e$-folds $N=60$. If the form of the  spectral index $n_s$ or other slow-roll parameters is given,  we can  reconstruct the inflationary potentials in the slow-roll regime \cite{Mukhanov:2013tua,Roest:2013fha,Garcia-Bellido:2014gna,Barranco:2014ira,
 Boubekeur:2014xva,Chiba:2015zpa,Creminelli:2014nqa,
 Gobbetti:2015cya,Lin:2015fqa,Fei:2017fub,Gao:2017uja,Gao:2017owg,Fei:2020jab}.
 In this way,  a wide range of inflation models
 that consistent with the observational data can be found.

There is an attractor  phenomenon in which a lot of inflationary models make the same prediction.  For example, the $R^2$ inflation \cite{Starobinsky:1980te} and the Higgs inflation
with the nonminimal coupling $\xi\phi^2 R$  \cite{Kaiser:1994vs,Bezrukov:2007ep} make the same prediction  of $n_s=1-2/N$ and $r=12/N^2$.
This prediction, then known as the  universal attractor,
was also derived from a more general inflation model with
the nonminimal coupling $\Omega(\phi)=[1+\xi f(\phi)]R$ and potential $\lambda^2 f^2(\phi)$
for an arbitrary function $f(\phi)$ in the strong coupling limit \cite{Kallosh:2013tua}.
This general inflation model is reduced to the Higgs inflation
by choosing   $f(\phi)=\phi^2$.
If the nonminimal coupling is $\Omega(\phi)=\xi f(\phi)R$,
then we obtain  the induced inflation,
which also predicts the universal attractor in the strong coupling limit \cite{Giudice:2014toa}.
For the two fields conformal invariant inflation model  with the  potential $F(\phi/\chi)(\phi^2-\chi^2)^2/36$,
under the gauge $\chi^2-\phi^2=6$ and in terms of the canonical scalar $\varphi$, the potential becomes $F(\tanh\varphi)$ \cite{Kallosh:2013hoa}. For the monomial function of $F$,
one gets the T-model  $V(\varphi)=V_0\tanh^{2n}(\varphi/\sqrt{6})$, that  also gives the universal attractor \cite{Kallosh:2013hoa}.
If the gauge is chosen as $\chi=\sqrt{6}$ instead, the kinetic term for the scalar field $\phi$ in the Einstein frame becomes $(\partial\phi)^2/(1-\phi^2/6)^2$ \cite{Kallosh:2014rga} and has a pole of order 2. In terms of the
canonical scalar $\varphi=\sqrt{6}\tanh^{-1}(\phi/\sqrt{6})$, the potential is also the T-model \cite{Kallosh:2013maa,Kallosh:2013daa} and the universal attractor is obtained.
The universal attractor can be generalized to the $\alpha$ attractors with the same $n_s$ and  $r=12\alpha/N^2$ by varying the K\"{a}hler curvature \cite{Kallosh:2013yoa}. 
For the $\alpha$ attractors,   the kinetic term is generalized to $\partial\phi^2/(1-\phi^2/6\alpha)^2$ in the  Einstein frame \cite{Kallosh:2014rga}, and the potential is generalized to $V(\varphi)=V_0\tanh^{2n}(\varphi/\sqrt{6\alpha})$ in terms of the canonical scalar field  \cite{Kallosh:2016gqp}. For more about the $\alpha$-attractors, see Refs. \cite{Linde:2015uga,Kallosh:2015lwa,Odintsov:2016vzz,Kumar:2015mfa,
Krajewski:2018moi,Yi:2018gse,Sabir:2019wel,AresteSalo:2021wgb,Shojaee:2021zmf,Aldabergenov:2020bsl}.
The common denominator of these  attractor inflation models was then elucidated in Ref. \cite{Galante:2014ifa} that their robust predictions stem from a joint pole of order 2  in the kinetic term of the inflaton field in the Einstein frame formulation prior to switching to the canonical variables.
The inflation model with a pole in  the kinetic term was discussed in detail as the
generalized pole inflation \cite{Broy:2015qna,Terada:2016nqg}.
It was then found that any attractor is possible  for the nonminimal coupling $\Omega(\phi)=[1+\xi f(\phi)]R$ with potential $\Omega^2(\phi)U(\sqrt{3/2}\ln\Omega(\phi))$ in the strong coupling limit \cite{Yi:2016jqr}.

Considering the importance of the pole in the kinetic term in the Einstein frame on predicting the attractors, in this paper, we  research the attractor phenomenon in the inflation model with a general non-canonical term.
We find that any  attractor is possible for this inflation model. We also show that
the $\alpha$ attractors can be obtained from a general non-canonical  kinetic term with a pole of order 2 in the Einstein frame for an arbitrary potential. The  inflation model  with a non-canonical kinetic term can be also used to   produce primordial black holes and induce secondary  gravitational waves \cite{Lin:2020goi,Yi:2020kmq,Yi:2020cut,Gao:2020tsa,Zhang:2020uek,Yi:2021lxc,Gao:2021vxb}.
The paper is organized as follows.
In Sec. II, we derive the attractor formulas in the inflation model with a non-canonical  kinetic term. The examples for the models with different attractors are given in Sec. III. The conclusions are drawn in Sec. IV.

\section{The attractors}
In this paper, we work on the attractor behavior in the inflation model with a non-canonical term in the Einstein frame. The action  is
\begin{equation}\label{gen:act1}
  S=\int d x^4 \sqrt{-g}\left[\frac{1}{2}R-\frac{1}{2}K(\phi)(\partial\phi)^2-V(\phi)\right],
\end{equation}
where $(\partial \phi)^2=g^{\mu\nu}\nabla_\mu\phi\nabla_\nu\phi $, $V(\phi)$ is the potential, and $K(\phi)>0$ is an arbitrary coupling function, and  we take the convention $8\pi G=1$. With the help of this non-canonical coupling function $K(\phi)$, the power spectrum of the primordial curvature perturbations can be enhanced at  small scales to produce primordial black holes and induce secondary gravitational wave, and simultaneously keep small  at large scales  to satisfy the constraints from the CMB  anisotropy \cite{Lin:2020goi,Yi:2020kmq,Yi:2020cut,Gao:2020tsa,Zhang:2020uek,Yi:2021lxc,Gao:2021vxb}.
In terms of the canonical field $\varphi$,
\begin{gather}\label{tran1}
  d\varphi =\sqrt{K(\phi)}d\phi,
\end{gather}
the model \eqref{gen:act1}  becomes
\begin{equation}\label{gen:act2}
   S=\int d x^4\sqrt{-g}\left[\frac{1}{2}R-\frac{1}{2}\nabla_\mu\varphi\nabla^\mu\varphi-U(\varphi)\right],
\end{equation}
where the canonical potential related to potential $V(\phi)$ is
\begin{gather}\label{tran1:pot}
 U[\varphi(\phi)]=V(\phi).
\end{gather}
By combining the relations \eqref{tran1} and \eqref{tran1:pot},
the coupling function $K(\phi)$  expressed by the potentials becomes
\begin{equation}\label{gen:kp}
K(\phi)=\left[\tilde{U}' (V(\phi))V'(\phi)\right]^2,
\end{equation}
where $\tilde{U}=U^{-1}$ is the inverse function of the canonical potential $U(\varphi)$,  and a prime denotes the derivative with respect to the argument of the function,
\begin{equation}
\tilde{U}' (V(\phi))=\frac{d \tilde{U}(x)}{d x}\bigg| _{x=V(\phi)~~},\quad V'(\phi)=\frac{d V(\phi)}{d \phi}.
\end{equation}
By using the relation \eqref{gen:kp}, the  action  \eqref{gen:act1}  becomes
\begin{equation}\label{gen:act3}
  S=\int d x^4 \sqrt{-g}\left[\frac{1}{2}R-\frac{1}{2}
  \left[\tilde{U}' (V(\phi))V'(\phi)\right]^2(\partial \phi)^2 -V(\phi)\right].
\end{equation}
Action  \eqref{gen:act2} and action \eqref{gen:act3} are equivalent and make the same prediction for $n_s$ and $r$.
Because the prediction of action \eqref{gen:act2} is determined by the canonical potential $U(\varphi)$, the prediction  of action \eqref{gen:act3} will be also determined by the potential $U(\varphi)$ and independent on the potential $V(\phi)$.
Therefore, for different potentials $V(\phi)$, the action \eqref{gen:act3} will make the same prediction, which is an attractor and determined by  the canonical  potential $U(\varphi)$. In the following, we will show explicitly how to obtain the attractors determined by the  T-model, the E-model, and the Hilltop potential.

\section{Examples}
\subsection{ $\alpha$ -attractors}
We first discuss how to obtain the $\alpha$-attractors from the model \eqref{gen:act3}. The $\alpha$-attractors are \cite{Kallosh:2013yoa}
\begin{equation}\label{attractors:nsr}
  n_s=1-\frac{2}{N},\quad r=\frac{12\alpha}{N^2},
\end{equation}
and consistent well with the Planck 2018 constraints \cite{Akrami:2018odb},
\begin{equation}\label{cmb:obser}
n_s=0.9649\pm 0.0042,\quad r_{0.002}<0.064.
\end{equation}
The canonical potentials that make the predictions of the $\alpha$ attractors are T-model and E-model.

\subsubsection{T-model}
The T-model  is
\begin{equation}\label{tmodel}
U(\varphi)=U_0\text{tanh}^{2n}\left(\frac{\varphi}{\sqrt{6\alpha}}\right),
\end{equation}
with $\alpha>0$ and $n>0$.
Substituting  T-model  \eqref{tmodel} into Eq. \eqref{gen:kp},
we obtain
\begin{equation}\label{tmodel:gen:k}
K(\phi)=\frac{3\alpha }{2n^2}\left(\frac{1}{U_0}\right)^{1/n}\frac{V^{\frac{1}{n}-2}}
{\left[1-\big(V/U_0\big)^{1/n}\right]^2}\left[\frac{dV(\phi)}{d\phi}\right]^2.
\end{equation}
For different potentials $V(\phi)$, the model \eqref{gen:act1} combined
with Eq. \eqref{tmodel:gen:k} will give the same prediction \eqref{attractors:nsr}.
The coupling function $K(\phi)$ has a pole of order $2$,
which is consistent with the statement in   Ref. \cite{Galante:2014ifa},
that all the cosmological attractors can be brought to the inflation model \eqref{gen:act1}
with the coupling function $K(\phi)$ having  a pole of order $2$.
Therefore, we obtain the $\alpha$ attractors from the general non-canonical term \eqref{tmodel:gen:k} with
a pole of order 2 for an arbitrary potential. To show the T-model \eqref{tmodel:gen:k} more specifically,
in the following, we choose the potential $V(\phi)$ as the  chaotic inflation potential or the natural inflation potential.

For the chaotic inflation, the  potential is \cite{Linde:1983gd}
\begin{equation}\label{chaotic}
    V(\phi)=V_0\phi^p,
\end{equation}
the corresponding coupling function Eq. \eqref{tmodel:gen:k} becomes
\begin{equation}\label{tmodel:chaotic:k1}
   K(\phi)= \frac{3 p^2\alpha }{2\beta n^2} \frac{\phi^{\frac{p}{n}-2}}{\left(1-\beta^{-1}\phi^{p/n}\right)^2},
\end{equation}
with $\beta=(U_0/V_0)^{1/n}$; and the action \eqref{gen:act3} becomes
\begin{equation}\label{tmodel:chaotic:act1}
  S=\int d x^4 \sqrt{-g}\left[\frac{1}{2}R-
   \frac{3 p^2\alpha }{2\beta n^2} \frac{\phi^{\frac{p}{n}-2}}{\left(1-\beta^{-1}\phi^{p/n}\right)^2}\frac{(\partial \phi)^2}{2}-V_0\phi^p\right],
\end{equation}
which is the general form of the chaotic inflation that gives the T-model.

To make the non-canonical term  reduces to the canonical case after inflation
at the low energy regimes $\phi\ll1$, we choose
\begin{equation}\label{tmodel:chaotic:1}
p=2n, \quad \beta=6\alpha.
\end{equation}
And then the coupling function   \eqref{tmodel:chaotic:k1} becomes
\begin{equation}\label{tmodel:chaotic:kk}
K(\phi)= \frac{1}{\left[1-\phi^{2}/(6\alpha)\right]^2},
\end{equation}
the attractor model \eqref{tmodel:chaotic:act1} becomes \cite{Kallosh:2013yoa,Kallosh:2014rga}
\begin{equation}\label{gen:tmodel1}
  S=\int d x^4 \sqrt{-g}\left[\frac{1}{2}R-\frac{1}{\left[1-\phi^{2}/(6\alpha)\right]^2}
  \frac{(\partial\phi)^2}{2}-V_0\phi^{2n}\right].
\end{equation}
Therefore, we obtain the $\alpha$-attractors action derived in Ref. \cite{Kallosh:2013yoa,Kallosh:2014rga}. And from the above discussion, we know that this action, reducing to the canonical model at the lower energy
regimes  $\phi\ll1$,  is a special case of action \eqref{tmodel:chaotic:act1}.

For the natural inflation, the potential is \cite{Freese:1990rb}
\begin{equation}\label{nat:vk}
  V(\phi)=V_0\left[1+\cos\left(\phi/f\right)\right],
\end{equation}
the corresponding coupling function Eq. \eqref{tmodel:gen:k} becomes
\begin{equation}\label{nat:tk}
  K(\phi)=\frac{3\alpha}{2n^2f^2\beta_2}
  \frac{\cos^{\frac{2}{n}-2}(\phi/2f)\sin^2(\phi/2f)}{\left[1- \beta_2^{-1}\cos^{2/n}(\phi/2f)\right]^2},
\end{equation}
with $\beta_2=\beta/2^{1/n}$. Substituting relations \eqref{nat:tk} and potential \eqref{nat:vk} into action \eqref{gen:act1}, we obtain the general natural inflation model that gives the T-model.
If we choose
\begin{equation}
\beta_2=1,\quad n=1,\quad f=\sqrt{3\alpha},
\end{equation}
the equation \eqref{nat:tk} becomes
\begin{equation}\label{gen:tmodel:nat:k}
    K(\phi)= \frac{1}{\big[1-\cos(\phi/\sqrt{3\alpha})\big]},
\end{equation}
and the attractor model   \eqref{tmodel:chaotic:act1} becomes
\begin{equation}\label{gen:tmodel:nat1}
  S=\int d x^4 \sqrt{-g}\left[\frac{1}{2}R- \frac{1}{\big[1-\cos(\phi/\sqrt{3\alpha})\big]}
  \frac{(\partial\phi)^2}{2}-V_0\left[1+\cos\left(\phi/\sqrt{3\alpha}\right)\right]\right].
\end{equation}
Therefore, we obtain the T-model from the natural inflation. 

\subsubsection{E-model}
The  E-model is
\begin{equation}\label{emodel}
U(\varphi)=U_0\left[1-\exp\left(-\sqrt{\frac{2}{3\alpha}}\varphi\right)\right]^{2n}.
\end{equation}
Substituting it into Eq. \eqref{gen:kp},  the corresponding coupling function becomes
\begin{equation}\label{emodel:gen:k}
K(\phi)= \frac{3\alpha}{8n^2}\left(\frac{1}{U_0}\right)^{1/n}  \frac{V^{\frac{1}{n}-2}}{\left[1-\big(V/U_0\big)^{1/2n}\right]^2}\left[\frac{dV(\phi)}{d\phi}\right]^2.
\end{equation}
The  E-model and the  T-model both predict the $\alpha$-attractors \eqref{attractors:nsr}. Therefore, the non-canonical coupling function for the E-model and that for the T-model are almost the same, except that the point of the pole in E-model is $\big(V/U_0\big)^{1/2n}$ while that in T-model  is  $\big(V/U_0\big)^{1/n}$.
To show the E-model \eqref{emodel:gen:k} more specifically,
 we also choose the potential $V(\phi)$ as the  chaotic inflation potential or the natural inflation potential.

For  the chaotic inflation with the potential \eqref{chaotic}, the non-canonical kinetic term becomes
\begin{equation}\label{emodel:chao:k}
    K(\phi)=\frac{3 p^2\alpha }{8\beta n^2} \frac{\phi^{\frac{p}{n}-2}}{\left(1-\phi^{p/(2n)}/\sqrt{\beta}\right)^2},
\end{equation}
and the action \eqref{gen:act3} becomes
\begin{equation}\label{emodel:chaotic:act1}
  S=\int d x^4 \sqrt{-g}\left[\frac{1}{2}R-
 \frac{3 p^2\alpha }{8\beta n^2} \frac{\phi^{\frac{p}{n}-2}}{\left(1-\phi^{p/(2n)}/\sqrt{\beta}\right)^2}\frac{(\partial \phi)^2}{2}-V_0\phi^p\right].
\end{equation}
This is  the general form of the chaotic inflation that gives the E-model. To get a canonical kinetic term after inflation at  the low energy regime $\phi\ll1$, we choose
\begin{equation}\label{emodel:chaotic:p1}
   p=2n, \quad \beta=\frac{3\alpha}{2}.
\end{equation}
And then the  coupling function becomes
\begin{equation}\label{emodel:chaotic:k1}
    K(\phi)= \frac{1}{\left(1-\sqrt{\frac{2}{3\alpha}}\phi\right)^2},
\end{equation}
the attractor action  \eqref{emodel:chaotic:act1} becomes
\begin{equation}\label{gen:tmodel2}
  S=\int d x^4 \sqrt{-g}\left[\frac{1}{2}R-\frac{1}{\left(1-\sqrt{\frac{2}{3\alpha}}\phi\right)^2}\frac{(\partial\phi)^2}{2}-V_0\phi^{2n}\right].
\end{equation}
In Ref. \cite{Kallosh:2014rga}, the E-model is obtained from the non-canonical kinetic coupling function \eqref{tmodel:chaotic:kk} with the potential
 $\left[\phi/(\sqrt{6\alpha}+\phi )\right]^{2n}$.
In this paper, we show that the  E-model can be also obtained  from the chaotic inflation with
a non-canonical term,   that reduces to the canonical case at the lower energy regimes $\phi\ll1$.

For the natural inflation with potential \eqref{nat:vk}, the coupling function reduces to
\begin{equation}\label{emodel:nat:k}
K(\phi)=\frac{3\alpha}{8n^2 \beta_2 f^2}
\frac{\cos^{\frac{2}{n}-2}(\phi/2f)\sin^2(\phi/2f)}{\left[1-\beta_2^{-1/2} \cos^{1/n}(\phi/2f) \right]^2}.
\end{equation}
If we choose
\begin{equation}\label{emodel:nat:parn}
  n=\frac{1}{2},\quad \beta_2=1,\quad f=\sqrt{\frac{3\alpha}{2}},
\end{equation}
 the coupling function becomes
\begin{equation}\label{emodel:nat:k2}
K(\phi)=
\frac{1+\cos\left(\sqrt{\frac{2}{3\alpha}}\phi\right)}{1-\cos\left(\sqrt{\frac{2}{3\alpha}}\phi\right)},
\end{equation}
and the action becomes
\begin{equation}\label{gen:starobinsky:nat1}
  S=\int d x^4 \sqrt{-g}\left[\frac{1}{2}R-
  \frac{1+\cos\left(\sqrt{\frac{2}{3\alpha}}\phi\right)}{1-\cos\left(\sqrt{\frac{2}{3\alpha}}\phi\right)}
  \frac{(\partial\phi)^2}{2}-V_0\left[1+\cos\left(\sqrt{\frac{2}{3\alpha}}\phi\right)\right]\right].
\end{equation}
Therefore, the E-model   can be also from the natural inflation.

\subsection{Hilltop inflation}
The other observational data favored model is the Hilltop inflation.
The  potential of the Hilltop inflation is \cite{ Boubekeur:2005zm}
\begin{equation}\label{hilltop:p}
    U(\varphi)=U_0\left[1-\left(\frac{\varphi}{\mu}\right)^n+\cdots\right].
\end{equation}
where the dots indicates higher-order terms. Because the Hilltop inflation is
supposed to take place near a maximum of the potential, the inflation field
satisfies $\phi/\mu\ll1$ and the higher-order terms can be neglected.
To achieve the small field inflation, we take $\mu<1$,
 and then the predictions for the spectral index  and  the tensor-to-scalar ratio are \cite{Yi:2016jqr}
\begin{equation}\label{hill:ns}
  n_s-1=-\frac{2(n-1)}{(n-2)N},
\end{equation}
\begin{equation}\label{hill:r}
  r=\frac{8n^2}{\mu^2}\left[\frac{\mu^2}{n(n-2)N}\right]^{(2n-2)/(n-2)},
\end{equation}
where we suppose  $n>2$.
Taking the $e$-folds  $N=60$, we obtain
 the  observational constraints \eqref{cmb:obser} on the power index of the Hilltop inflation,
\begin{equation}\label{constraint:n}
  n>7.59.
\end{equation}
Combining Eq. \eqref{gen:kp} and the Hilltop inflation potential \eqref{hilltop:p}, we obtain the non-canonical term with the Hilltop inflation attractors,
\begin{equation}\label{hilltop:k}
K(\phi)=\frac{\mu^2}{n^2 U_0^2}\frac{1}{\left[1-\left(\frac{V}{U_0}\right)\right]^{2-\frac{2}{n}}}\left[\frac{dV(\phi)}{d\phi}\right]^2.
\end{equation}
From  condition \eqref{constraint:n},  we have
\begin{equation}\label{hilltop:pole}
  1.74<2-\frac{2}{n}<2.
\end{equation}
Therefore, the  non-canonical coupling for the Hilltop attractor has  a pole of order  from $1.74$ to $2$.

For  the  chaotic inflation \eqref{chaotic}, the non-canonical coupling function \eqref{hilltop:k} becomes
\begin{equation}\label{hilltop:chao:k}
 K(\phi)= \frac{\mu^2 p^2}{\gamma n^2} \frac{\phi^{2p-2}}{\big(1-\gamma^{-1/2}\phi^p\big)^{2-\frac{2}{n}}},
\end{equation}
where $\gamma=(U_0/V_0)^2$. If we want the non-canonical term to reduce to canonical term after inflation at the low energy  regime  $\phi\ll1$, we  choose
\begin{equation}\label{hilltop:chao:p}
  p=1,\quad \gamma=\frac{\mu^2}{ n^2},
\end{equation}
and the corresponding coupling function is
\begin{equation} \label{hilltop:chao:k1}
  K(\phi)= \frac{1}{\big(1-n \phi/\mu\big)^{2-\frac{2}{n}}},
\end{equation}
and the action becomes
\begin{equation}\label{hilltop:chao:act1}
  S=\int d x^4 \sqrt{-g}\left[\frac{1}{2}R-
 \frac{1}{\big(1-n \phi/\mu\big)^{2-\frac{2}{n}}}\frac{(\partial\phi)^2}{2}
  -V_0\phi \right].
\end{equation}
Therefore, we obtain the Hilltop inflation  from the chaotic inflation model. 
This action is similar to the action of the E-model \eqref{gen:tmodel2}, 
except that the poles in each kinetic term are different.

For the natural inflation \eqref{nat:vk}, the  non-canonical term \eqref{hilltop:k} becomes
\begin{equation}\label{hilltop:nat:k}
  K(\phi)=\frac{\mu^2}{f^2n^2\gamma_2}\frac{\cos^2(\phi/2f)\sin^2(\phi/2f)}
  {\left[1-\cos^2(\phi/2f)/\sqrt{\gamma_2}\right]^{2-\frac{2}{n}}},
\end{equation}
where $\gamma_2=\gamma/4$. Choosing
\begin{equation}
\gamma_2=1,\quad f=\frac{\mu}{2^{1/n}n},
\end{equation}
the non-canonical coupling term becomes
\begin{equation}\label{hilltop:nat:k1}
  K(\phi)= \frac{1+\cos( 2^{1/n}n\phi/\mu)}
  {\left[1-\cos( 2^{1/n}n\phi/\mu)\right]^{1-\frac{2}{n}}},
\end{equation}
and the action becomes
\begin{equation}\label{hilltop:nat:act1}
  S=\int d x^4 \sqrt{-g}\left[\frac{1}{2}R-
 \frac{1+\cos( 2^{1/n}n\phi/\mu)}
  {\left[1-\cos( 2^{1/n}n\phi/\mu)\right]^{1-\frac{2}{n}}}
  \frac{(\partial\phi)^2}{2}-V_0\left[1+\cos( 2^{1/n}n\phi/\mu)\right]\right].
\end{equation}
Therefore, we obtain the Hilltop inflation from the natural  inflation model.
\section{Conclusion}
There exists a  phenomenon of attractor  that many inflationary models  make the same prediction for $n_s$ and $r$.  The inflation model   with the non-canonical kinetic  $K(\phi)=[\tilde{U}'(V(\phi))V'(\phi)]^2$ and  an arbitrary potential  $V(\phi)$ is equal to the canonical inflation model with the canonical potential $U(\varphi)$.
Therefore, for different inflation potentials $V(\phi)$, the models with that $K(\phi)$   make  the same prediction which is determined by the canonical potential $U(\varphi)$. Thus, by reconstructing the potential from the  observable $n_s$ or $r$, we can get any attractor in the inflation models with the non-canonical kinetic.
In particular,  we show explicitly the models that give the attractors determined  by the T model,  the E model, and the  Hilltop potential, respectively.
For an arbitrary potential, the kinetic terms  $K(\phi)$ in these models all have a pole.  The order of the pole determined by the T-model and the E-model is $2$ and that determined by  the Hilltop potential is $2-2/n$.

In conclusion, any attractor is possible for the inflation model with a non-canonical term, and the presence of attractors poses a challenge to differentiate inflation models.

\acknowledgments
This work was supported by the National Natural Science Foundation of China under Grants
Nos. 11633001, 11920101003 and 12021003,
the Strategic Priority Research Program of the Chinese Academy of Sciences,
Grant No. XDB23000000 and the Interdiscipline Research Funds of Beijing Normal University.
Z. Y. was supported by China Postdoctoral Science Foundation Funded Project under Grant No. 2019M660514.



\providecommand{\href}[2]{#2}\begingroup\raggedright\endgroup

\end{document}